# Out-of-Plane Biphilic Surface Structuring for Enhanced Capillary-Driven Dropwise Condensation

*Luca Stendardo[1,†], Athanasios Milionis[1], George Kokkoris[2], Christos Stamatopoulos[1], Chander Shekhar Sharma[3], Raushan Kumar[3], Matteo Donati[1], Dimos Poulikakos[1,*]*

[1]Laboratory of Thermodynamics in Emerging Technologies (LTNT), Sonneggstrasse 3, ETH Zurich, 8092 Zurich, Switzerland

[2]Institute of Nanoscience and Nanotechnology, NCSR Demokritos, Agia Paraskevi 15341, Greece

[3]Thermofluidics Research Lab, Department of Mechanical Engineering, Indian Institute of Technology Ropar, Rupnagar, Punjab, 140001 India

**Keywords:** Condensation, Biphilic, Micro-pyramids, Droplet jumping, Capillary-driven, Volume of Fluid method

ABSTRACT

Rapid and sustained condensate droplet departure from a surface is key towards achieving high heat transfer rates in condensation, a physical process critical to a broad range of industrial and societal applications. Despite progress in enhancing condensation heat transfer through inducing its dropwise mode with hydrophobic materials, sophisticated surface engineering methods that can lead to further enhancement of heat transfer are still highly desirable. Here, by employing a three-dimensional, multiphase computational approach, we present an effective out-of-plane biphilic surface topography, that reveals an unexplored capillarity-driven departure mechanism of condensate droplets. This texture consists of biphilic diverging micro-cavities wherein a matrix of small hydrophilic spots is placed at their bottom, that is, amongst the pyramid-shaped, superhydrophobic micro-textures forming the cavities. We show that an optimal combination of the hydrophilic spots and the angles of the pyramidal structures can achieve high deformational stretching of the droplets, eventually realizing an impressive "slingshot-like" droplet ejection process from the texture. Such a droplet departure mechanism has the potential to reduce the droplet ejection volume and thus enhance the overall condensation efficiency, compared to coalescence-initiated droplet jumping from other state-of-the-art surfaces. Simulations have shown that optimal pyramid-shaped biphilic micro-structures can provoke droplet self-ejection at low volumes, up to 56% lower compared to superhydrophobic straight pillars, revealing a promising new surface micro-texture design strategy towards enhancing condensation heat transfer efficiency and water harvesting capabilities.

INTRODUCTION

Condensation is a necessary step of the natural water cycle, but it is also of fundamental importance to the energy sector, e.g., in thermal power generation[1] and thermal management of micro-processors,[2] as phase change can drastically increase heat transfer. Traditionally, the functionality of industrial condensers involves condensation on metallic tubes made from aluminum or copper which are inherently hydrophilic and thus promote filmwise condensation, limiting the heat transfer efficiency. On the other hand, hydrophobic surfaces usually enable dropwise condensation wherein water droplets nucleate, grow, coalesce, and depart periodically. This leads to a significantly higher overall heat transfer coefficient.[3,4] For this reason, several works in recent years have investigated surface characteristics enabling passive, controlled droplet departure during dropwise condensation on metallic surfaces.[5–7]

Texturing a hydrophobic surface can result in superhydrophobicity which is characterized by ultralow contact angle hysteresis (typically less than 10°) and, therefore, small droplet departure diameters. This can only be achieved with the presence of a hydrophobic micro- / nano-texture.[8] Although the nano-scale texture is critical for superhydrophobicity during condensation to efficiently remove the small condensate droplets, the micro-scale texture can be added to synergistically assist droplet departure through the generation of Laplace pressure imbalance.[8–10] This principle can be utilized to generate passive wetting transitions of the condensed droplets from Wenzel to Cassie state at the length scale of the surface texture[10] through rational design of the micro-topography. Superhydrophobicity at nano-scale also enables droplet departure through coalescence-induced droplet jumping (CIDJ) during condensation.[6] This is a gravity-independent droplet departure mechanism which allows passive and rapid removal of condensed water to enhance anti-icing,[11] defrosting,[12] and self-cleaning[13] properties of surfaces. The low adhesion force on superhydrophobic surfaces allows the conversion of a larger amount of excess surface energy into kinetic energy, due to total surface area reduction during coalescence, enabling jumping events. This phenomenon can potentially lead to superior condensation heat transfer[5,6,14] since it affects much smaller droplets compared to the size threshold requirement for the gravitational removal of droplets. Given that it is a rather random occurring phenomenon, research has been directed towards designing rational textures that tune the size, velocity, and departure direction of jumping droplets.[14–16]

A variety of micro-feature-geometries that can enable pressure gradients within the droplets and manipulate the droplet movement, such as micro-pillars or micro-cones, have already been experimentally investigated.[10,14,17–20] In particular, it has been shown that diverging micro-cavity geometries, defined by their half-opening angle β, enhance droplet-ejection by creating a favorable Laplace pressure imbalance and, additionally, increase the surface area available for vapor condensation and overall heat transfer.[9,17,18,21] Theoretical studies have shown that the highest Laplace pressure imbalance is achieved for β ≈ 7°.[9] Optimizing the micro-cavity geometry experimentally is challenging and not cost-effective due to the inherent nature of the fabrication process. Thus, simulating condensation over surfaces with different micro-cavity opening angles can give new insights on the ideal micro-scale structure, avoiding long and costly trial-and-error experimental parametric investigations.

Previous computational studies have been able to successfully show how the droplet growth behavior strongly depends on the number of nucleation sites,[22] as well as to derive a mathematical model that can predict individual droplet evolution, droplet coalescence, and droplet departure. Additionally, contact angle and hysteresis variations have also been included and the results of the simulations have been validated with experimental data.[23] Subsequently, detailed droplet dynamics and heat transfer performance of different wettability patterns have been numerically analyzed. In particular, the effects of a pillar micro-structure on droplet dynamics (especially droplet coalescence jump, pillar squeezing droplet jump, and droplet dragging by wettability gradients) were investigated.[24]

In this work, we go beyond analyzing well-known condensation modes and we explore sophisticated surface designs that can lead to unique condensation outcomes. Using a computational framework coupling a 3D volume of fluid (VOF) model with the continuity and momentum equations, we propose a novel strategy for enhanced condensate droplet jumping. It involves combining superhydrophobic divergent micro-structures in the form of micro-pyramids with local hydrophilic spots at the bottom of the micro-cavities the micro-pyramids form amongst them. We find that this out of plane biphilic texture can trigger enhanced jumping of individual condensate droplets due to a synergistic combination of the Laplace pressure imbalance induced by diverging cavity, and local pinning induced by the hydrophilic spots. This results in a "slingshot"-like self-ejection of droplets from the micro-texture at lower volumes compared to existing approaches for droplet removal from surfaces. Ultimately, these surface designs open new possibilities to tune droplet ejection during condensation which is critical for both heat transfer and water collection applications.

## RESULTS AND DISCUSSION

### Single Droplet Jumping on Biphilic Micro-Cavities

The working principle of the biphilic micro-cavities is presented in Figure 1. Our approach consists of investigating droplet jumping due to droplet growth in different regular surface micro-structures composed of micro-pillars and micro-pyramids (Fig. 1a). In the first (control) case, a superhydrophobic surface is applied (Fig. 1b), while in the second case we introduce a hydrophilic spot in the cavities surrounded by otherwise superhydrophobic micro-pyramids, resulting in out-of-plane biphilic surface structures (or biphilic cavities, Fig. 1c). We investigate the effect of surface textures towards removal of condensate droplets from surfaces by utilizing Laplace pressure imbalance and the addition of hydrophilic spots that induce the slingshot effect. To this end, we focus on the dynamics of individual condensate droplets as they grow in diverging superhydrophobic micro-cavities formed by arrays of micro-pillars or micro-pyramids. We simulate growth of a single condensate droplet in one unit cell of such a regular micro-texture and investigate the effect of texture geometry on droplet mobility. Five different texture geometries are simulated: pyramids with a half-opening angle β of 0° (pillars), 7°, 14°, 20°, and 27° respectively. The micro-elements have a constant height (H) of 56 μm and the pitch (P, center to center) distance is 81 μm.

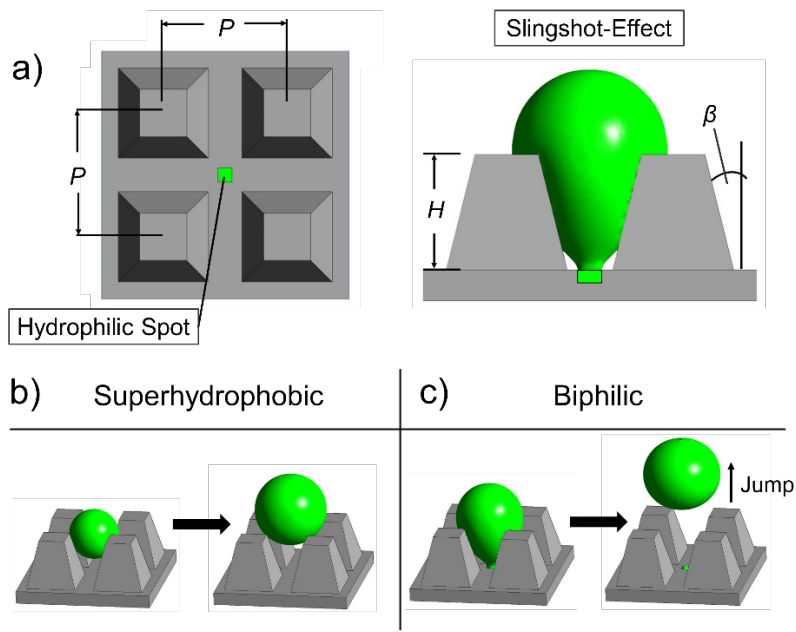

**Figure 1**. a) Cross section of the micro-elements. H = 56 μm and P = 81 μm are kept constant. The half-opening angle is denoted by the Greek letter *β*. Comparison between superhydrophobic (b) and biphilic (c) micro-cavities. The biphilic structures induce controlled droplet pinning at the bottom, enabling pressure gradients within the droplet. Detachment from the hydrophilic spot results in conversion of surface energy into kinetic energy, causing surface clearing droplet jumping.

The goal of this analysis is to investigate how individual droplets are affected by the micro-texture per se, excluding the influence of coalescence-induced droplet depinning. To simulate the growth of an individual micro-droplet in this geometry, a user-defined mass transfer model is used. The initial condition for this simulation consists of a small droplet (0.001 nL) placed at the base of the micro-cavity. Subsequently, the model "grows" the droplet through mass sources of liquid water in every cell of the droplet. The mass source is configured such that the droplet radius growth rate follows a power law and such that the rate reduces with increasing flow-time.[8]

The domain is discretized using a uniform mesh structure. Mesh resolution is chosen so as to allow the use of dynamic contact angle model[25] in order to account for the effect of nano-structuring on micro-structures. To simulate a superhydrophobic surface, the advancing contact angle $\theta_a$ is set to 167.8°, the receding $\theta_r$ to 165.9°, and the static $\theta$ to 166.9°. The contact angle hysteresis $\Delta\theta = \theta_a - \theta_r$ is therefore 1.9°. These values stem from experimental measurements on a perfluorodecanethiol functionalized copper hydroxide nano-structured surface.[26] Additionally, the hydrophilic spot has a size of 85 µm² and the contact angle on the spot is set to 20°.

Fig. 2a depicts the results from five different micro-structures [half-opening angles 0° (pillars), 7°, 14°, 20°, and 27°]. For the first set of simulations, the micro-features are simulated with a uniform superhydrophobic surface without hydrophilic spot, characterized by the aforementioned wetting parameters. For every texture geometry, a 4-frame sequence during the growth of a single droplet in the superhydrophobic micro-cavity is shown.

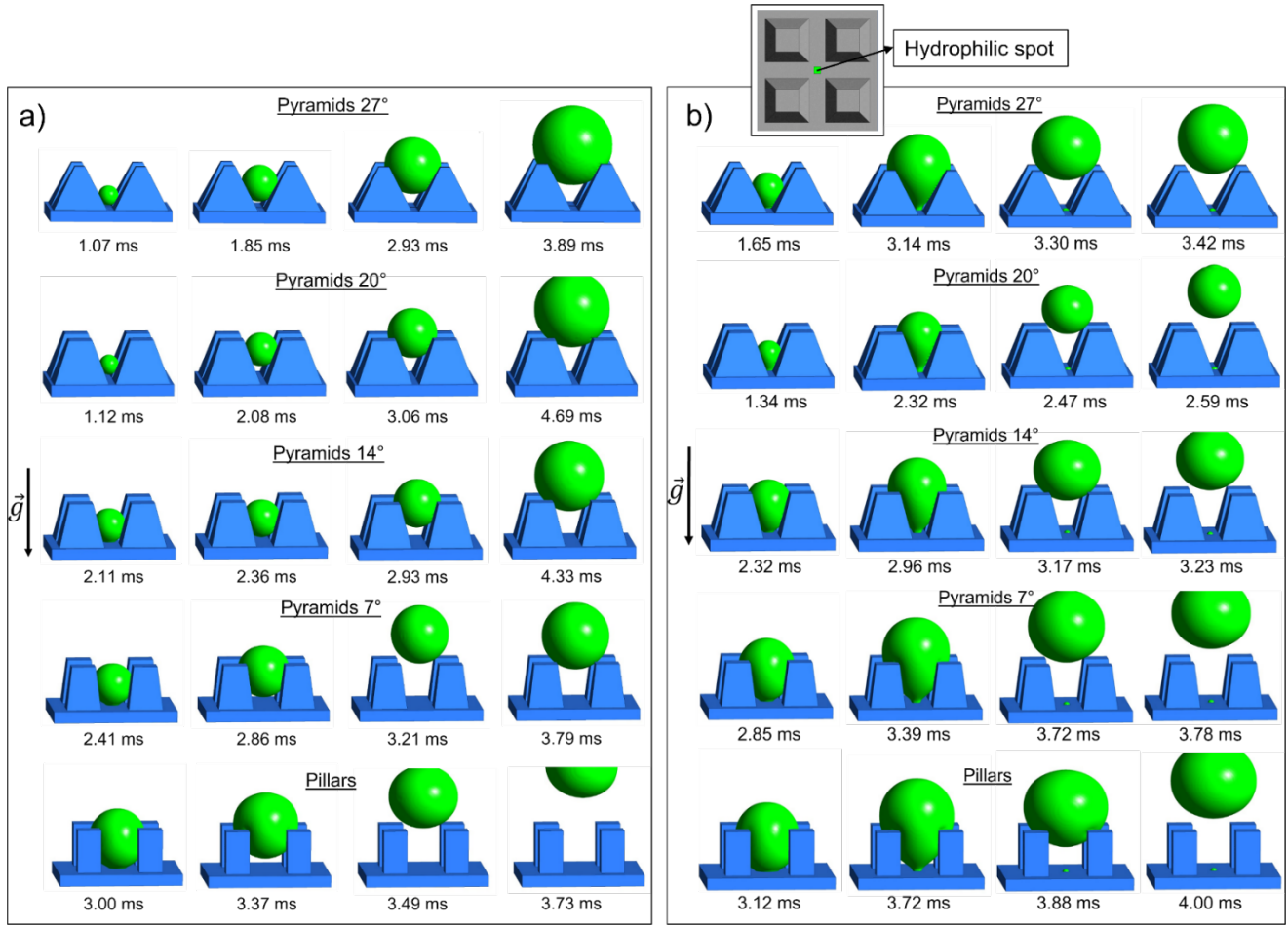

**Figure 2**. a) Single droplet growing at the center of the computational domains. Five different micro-geometries have been simulated, including uniform superhydrophobic surface in all cases with $\theta = 166.9°$, $\theta_r = 165.9°$, and $\theta_a = 167.8°$. The number following the text "Pyramids" or "Pillars" indicates the angle β. For the β = 7° case, the droplet barely clears the micro-texture, while for pillars, the droplet jumps out of the computational domain. b) The same as (a), but with a hydrophilic spot ($\theta = 20°$, A = 85 µm2) at the center of the domain. Surface clearing jumping events are observed in all cases. Inset figure illustrates top view of the texture along with location of the hydrophilic spot. Each 4-frame sequence shows the growth of the droplet, representing the detachment and/or ejection of the droplet.

During growth, the droplet is detached from the base of the micro-cavity as a consequence of the spatial constriction by the micro-elements. After detachment (due to low superhydrophobic surface adhesion), in all cases the droplets move upwards and out of the micro-cavity driven by the increase in volume and the action of Laplace pressure imbalance induced by the sidewall confinement. Droplet jumping is observed for the case of the pyramids with 7° inclined walls and the pillars. In particular, the 7° pyramids cause a small jump, enough to barely move the droplet out of the micro-cavity, but not with high enough velocity to completely remove the droplet from the computational domain. In fact, the droplet returns and sits on top of the micro-pyramids (gravity is acting against the droplet movement). The pillars, on the other hand, allow for a jump that completely removes the droplet from the surface. This can be explained by the fact that the vertical walls cause the droplet to remain inside the micro-cavity up to a much higher volume, and the pillars are able to exert a higher constriction and pressure on the droplet compared to the pyramids. Evidently, this results in a more pronounced squeezing of the droplet and a higher release of excess surface energy in the form of kinetic energy, therefore causing a jump with higher departure velocity. On the other hand, the shape of the pyramid sidewalls causes a Laplace pressure imbalance that drives the droplet gradually upwards, thus forcing it to stay less time inside the cavity and therefore accumulate less water mass compared to pillars. Eventually, this first set of simulations interestingly shows how vertical walls can induce a higher constriction on the droplet during the growing phase, compared to the inclined walls as in the pyramid case.

Going beyond the aforementioned findings, we introduce here a new approach that has the potential to further enhance the droplet ejection from diverging micro-geometries. Instead of a uniform superhydrophobic surface at the nano-scale, we introduce a hydrophilic spot at the center of the micro-cavity at its bottom (Fig. 2b). The hydrophilic spot, surrounded by the superhydrophobic background, can induce controlled droplet pinning at the droplet bottom, without altering the low adhesion due to high contact angle on the remaining contact area between the droplet and the micro-geometry. Therefore, during growth and for a certain time, the droplet can remain pinned to the hydrophilic spot, thereby delaying its detachment from the base of the micro-cavity. As a result, the pyramids can deform (stretch) the droplet to a greater extent, increasing the capillary pressure gradient inside the droplet to a critical point where the droplet "snaps" and detaches itself from the spot. The intensity of the pinning force can be adjusted by the size of the hydrophilic spot and/or $\theta$.

Following the above argument, another set of simulation runs is performed for biphilic micro-cavities consisting of the same five superhydrophobic ($\theta = 166.9°$) micro-geometries as shown in Fig. 2a, but with an additional hydrophilic spot (Fig. 2b). The effect of the hydrophilic spot on the droplet ejection behavior can be immediately observed especially for the pyramids with half-opening angle of 20° and 14°. The droplet is held inside the micro-cavity due to contact line pinning at the hydrophilic spot, allowing the droplet to deform. This deformation is kept up until the droplet finally snaps away from the hydrophilic spot. At that moment, the drop returns to a spherical shape and the released surface energy is converted into kinetic energy, thus allowing the droplet to jump suddenly in the vertical direction away from the surface, resembling a slingshot. Fig. 2b highlights how this new "slingshot effect" can cause surface clearing jumping events on all five micro-geometries and this is a key difference compared to the cases without a hydrophilic spot.

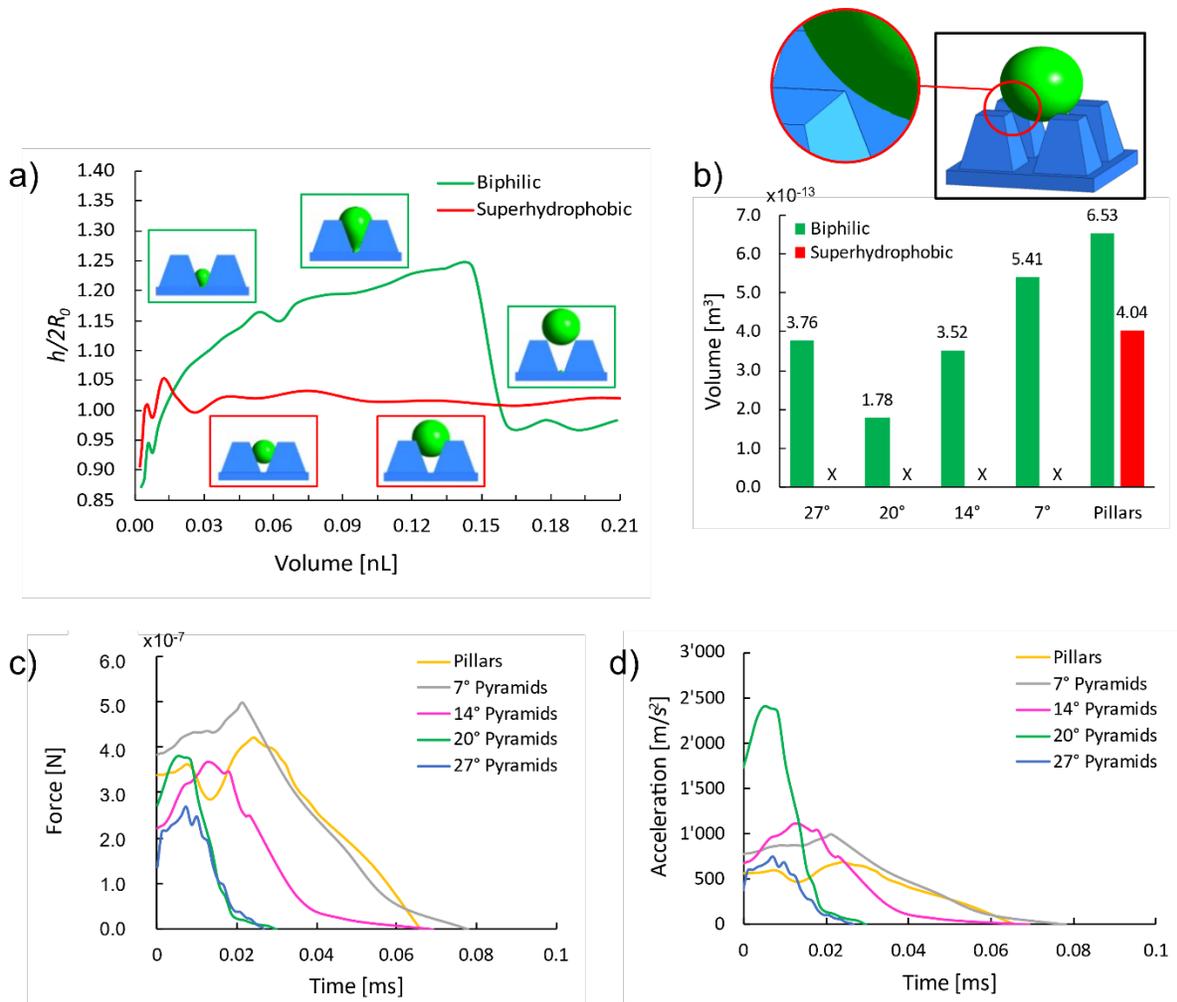

**Figure 3**. a) Calculated geometrical aspect ratio $h/(2R_0)$ as a function of the droplet volume for biphilic surface (green) and superhydrophobic surface (red). h represents the distance from the lowest to the highest point of the droplet, while $R_0$ represents the radius of a perfectly spherical droplet at the given volume. b) Volume at which the droplet clears the surface (loses contact) represented for five different micro-geometries with biphilic surface (green) and superhydrophobic surface (red). An "X" instead of a bar indicates that the droplet was not completely ejected from the surface. Inset figure illustrates the moment of detachment from the substrate, which was taken as ejection condition. c) Vertical force and (d) vertical acceleration of droplet, from the moment of detachment from the hydrophilic spot to complete lift off from the surface.

Fig. 3a presents a comparison of the geometrical droplet aspect ratio [$h/(2R_0)$, h is the distance from the lowest to the highest point of the droplet on y axis, $R_0$ is the radius of a perfectly spherical droplet at the given volume] for the 20° pyramids, with (green curve) and without (red curve) the hydrophilic spot. The results are plotted as a function of the droplet volume and show how the hydrophilic spot can increase the aspect ratio (thus induce larger deformation). At the moment of detachment from the hydrophilic spot, the aspect ratio is abruptly reduced, indicating the release of excess surface energy resulting in the droplet jump.

To further compare these two surfaces quantitatively, the surface clearing volumes (the volume at which the droplet loses contact with the micro-cavity surface) are calculated and compared in Fig. 3b. The biphilic surface gives a significant advantage on all diverging micro-geometries: the droplet is self-ejected due to the "slingshot effect". On the uniformly superhydrophobic surface (no hydrophilic spot) the droplet is gradually dragged out of the micro-cavity and ends up sitting on top of the micro-elements. Without the stretching effect of the hydrophilic spot, individual droplets are not able to clear the surfaces with diverging micro-geometry.

The 20° pyramids with hydrophilic spot achieve overall the smallest droplet ejection volume across all the geometries, 56% lower than the superhydrophobic straight pillars. This result indicates that the opening angle and hydrophilic spot can be optimally combined to reduce the droplet departure size through gravity-independent individual droplet jumping. This in turn can enable the highest frequency of condensation cycle among all geometries considered in this study. The introduction of a hydrophilic spot on superhydrophobic surfaces represents therefore a new approach to reducing the droplet departure volume, reducing the risk of surface flooding, and potentially increasing the condensation heat transfer efficiency.

To further characterize the surfaces equipped with a hydrophilic spot, an analysis on the droplet ejection force is carried out. Specifically, the reaction force from the substrate on the droplet is calculated, from the moment of detachment from the hydrophilic spot to the complete lift-off of the droplet from the surface. The magnitude of this reaction force reflects the sum of external forces acting on the droplet. In this case, only the vertical component of this reaction force is considered since the horizontal components are cancelled out due to symmetry. The reaction force from the surface reduces to zero at the moment the droplet loses contact with the micro-cavity walls (i.e., lifts-off). By dividing this vertical reaction force by the mass of the droplet, we obtain the vertical acceleration of the droplet. Figures 3c and 3d show both the vertical reaction force and vertical acceleration as a function of time. The graphs show that the more the pyramid walls are inclined (higher β), the quicker the force reduces to zero. This means that less time is elapsed from the moment of detachment from the spot to complete lift-off of the droplet from the surface. This can be seen for the cases of the 27° and 20° pyramids. The magnitude of the peak force is approximately the same for all cases. However, the size of the droplet at the moment of jumping is not the same for these cases: the droplet is smallest in the case of the 20° pyramids and is the largest in the case of the pillars. It can be seen that the 20° pyramids with hydrophilic spots exert by far the highest acceleration on the droplet (see Fig. 3d). From this point of view, the hydrophilic spot brings the highest benefit when the inclination angle of the micro-geometry is optimized: not only it can provoke droplet jumping for the geometries where droplet jumping is not observed otherwise (Fig. 2), but it can also maximize the droplet acceleration, enabling enhanced self-ejection.

Multiple Droplet Interaction

The previous analysis has shown how the hydrophilic spot increases the tendency for individual droplet jumping on all surfaces except for the pillars (Fig. 2b). In this section, an additional set of simulations on a 6-element micro-structure is performed. Here we are interested to see if using a larger surface, where more droplets are present and can interact with each other, can possibly affect the phenomena that were observed before for smaller domains where the focus was on single droplets. The simulations have been performed for the 20° pyramids, that showed the best results in the single droplet analysis, and for the straight pillars, where the surface with the hydrophilic spot showed a worse performance compared to the uniformly superhydrophobic surface (Fig. 3b).

In Fig. 4a two droplets are grown in neighboring unit cells of 20° micro-pyramids. Interestingly, the two droplets are self-ejected before they reach a sufficient size to coalesce. While coalescence-induced droplet jumping (CIDJ) is known to reduce the droplet removal size threshold compared to gravity driven techniques, the out-of-plane biphilic surfaces show the potential to reduce this size threshold even more compared to CIDJ. This confirms the capability to increase the frequency of condensation cycle compared to conventional superhydrophobic surfaces.

Fig. 4b shows two droplets growing in straight walled pillar biphilic cavities. The droplets are pinned to the hydrophilic spot, however when coalescence occurs, the released surface energy is large enough to induce depinning from both hydrophilic spots and trigger coalescence driven ejection of the droplets. The volume of the individual droplets right before coalescence is $2.97 \times 10^{-13}$ m$^3$, which is lower than the single droplet ejection volume on superhydrophobic pillar domain and biphilic pillar domain (see Fig. 3b) and closer to droplet ejection volume $1.78 \times 10^{-13}$ m$^3$ for micro-pyramids with $\beta = 20°$. Therefore, the seeming performance disadvantage of the biphilic pillar surface reduces significantly and it can be seen how droplet coalescence can work synergistically with the slingshot effect, when passive removal of droplets without the need for gravity is desired.

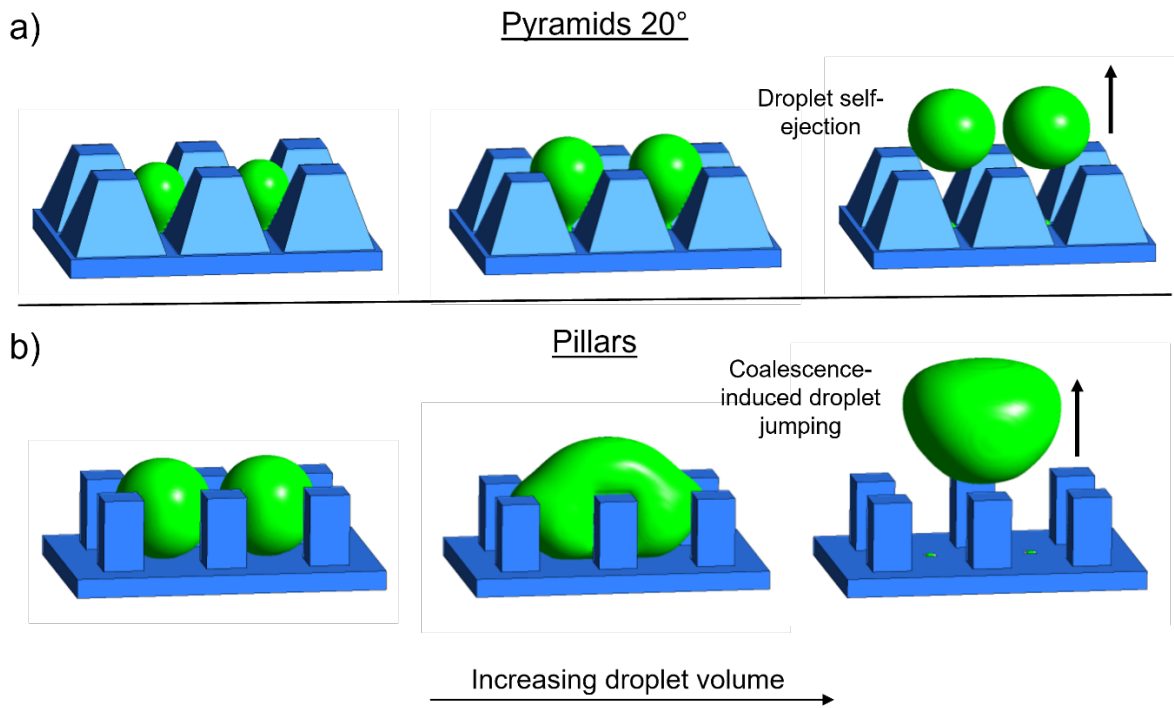

**Figure 4**. Comparison between two double-domains of (a) 20° pyramids and (b) straight-walled pillars, both equipped with hydrophilic spots. The three-frame sequence shows the effect of increasing droplet volume. In the case of the 20° pyramids, the droplets are stretched inside the micro-cavity, similar to what is observed on the single droplet domain. The droplets are self-ejected, without interacting with the neighboring droplet. On the pillar domain, coalescence with the neighbor droplet results in ejection at lower individual volume compared to what was observed on the single droplet domain.

In summary, the hydrophilic spot improves the droplet ejection behavior on all pyramid shaped micro-geometries: due to contact line pinning, the droplets are kept inside the micro-cavities, where they experience a higher degree of deformation compared to similar surfaces without the hydrophilic spot. Depinning through snapping away from the hydrophilic spot allows the droplets to regain a spherical shape, and the resulting release of surface energy induces surface clearing jumping events. The half-opening angle of 20° combined with a hydrophilic spot showed the smallest surface clearing volume, more than 49% lower than any other investigated surface geometry here. The droplets are self-ejected, reducing the jumping volume threshold compared to CIDJ.

We envisage that the rapidly developing micro-fabrication techniques can be explored towards realizing this biphilic texture for the improvement of condensation heat transfer efficiency in realistic applications.

COMPUTATIONAL METHOD

The results presented in this study are obtained using the commercial computational fluid dynamics (CFD) software Ansys 17.2/Fluent. The 3D volume of fluid (VOF) model is coupled to continuity and momentum equations. A user-defined mass transfer model is used for the interfacial mass transfer. A pressure-based finite volume solver is utilized, the SIMPLE-scheme is applied for the pressure-velocity coupling, and PRESTO! is used for pressure discretization. Second-order upwind schemes are used for the discretization of the momentum and energy equations, while the geometric reconstruction scheme from the work of Youngs[27] is used for interface calculations. Standard tolerances for convergence criteria are used, namely $10^{-3}$ for continuity and momentum, and $10^{-6}$ for energy conservation equation. A detailed representation of the computational domain is shown in Fig. 5. The same general structure of the domain is consistently utilized for all the simulations performed in this study.

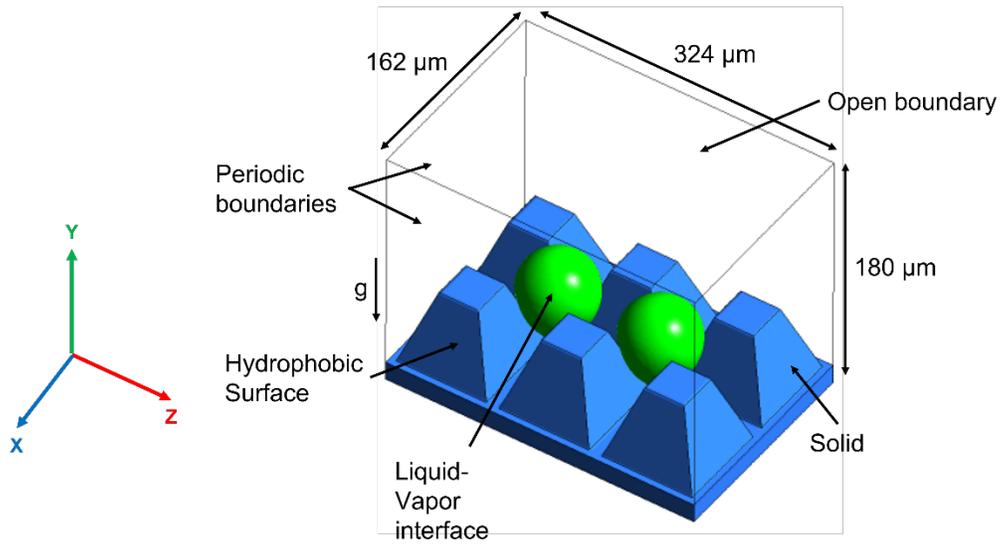

**Figure 5**. Geometry and boundary conditions. A rectangular computational domain composed of solid and fluid subdomains is used. The base material is shown in blue, while the liquid-vapor interface is depicted in green. Saturated steam represents the working fluid.

The domain is composed of one solid and two fluid subdomains (liquid water and saturated steam). The solid plate is textured with the micro-elements. The top surface (XZ-plane) of the computational domain is modeled as an open boundary. The four surrounding surfaces (XY and YZ-planes) are imposed as periodic boundaries. Throughout the various simulations, the initial conditions are imposed such that the fluid volume is composed of saturated water vapor with small droplet (0.001 nL) placed at the base of the micro-cavity. The contact angle of the liquid droplets at the solid/gas interface can be adjusted as desired through a dynamic contact angle model, which takes into account the contact angle hysteresis.

CONCLUSIONS

Through detailed simulations, we show that droplet removal through jumping can be enhanced by increasing and optimizing the confinement effect from the surrounding micro-structures. This ensures that the conversion of surface energy to kinetic (removal) energy is optimally utilized. We show that this effect can be exploited by combining a pyramid-shaped micro-structure, a superhydrophobic surface, and hydrophilic spots placed at the bottom, amongst the pyramid elements. Such a texture causes "slingshot"-like droplet ejection from the surface due to additional deformation on the droplet followed by a sudden "snapping" ejection event. In particular, the 20° pyramids combined with hydrophilic spots show the smallest droplet ejection volume across all the considered geometries. Compared to the superhydrophobic pillar structure, where self-ejection was observed, the ejection volume could be reduced by 56%.

A smaller droplet ejection volume increases the frequency of the condensation cycle improving the heat transfer efficacy. The introduction of hydrophilic spots on hierarchically structured superhydrophobic surfaces represents therefore a novel, unexplored approach to reducing the droplet departure volume and increasing the condensation efficiency. The concept of out-of-plane biphilic surfaces is relevant for enhancement of other processes that involve condensation as well, such as water harvesting. To this end, the improvement of droplet mobility can lead to larger amounts of condensate that leave the surface at potentially longer distances from the cooled surface due to the slingshot effect.

AUTHOR INFORMATION

**Corresponding Author**

*E-mail: dimos.poulikakos@ethz.ch

**Present Address**

†Department of Materials Science, University of Milano – Bicocca, via R. Cozzi 55, 20125 Milano, Italy

AUTHOR CONTRIBUTIONS

The manuscript was written through the contributions of all authors. All authors have given approval to the final version of the manuscript.

NOTES

The authors declare no competing financial interest.

ACKNOWLEDGMENTS

This project received funding from the European Union's Horizon 2020 research and innovation program under grant number 801229 ('HierARchical Multiscale NanoInterfaces for enhanced Condensation processes' - HARMoNIC).